\documentclass[12pt]{article}
\usepackage{graphicx}
\usepackage{epsf,amsmath,bbold,amsfonts,stmaryrd}

\usepackage[utf8]{inputenc}

\usepackage{mathrsfs}
\usepackage{appendix}
\usepackage{amssymb}
\usepackage{float}
\usepackage{color} 
\usepackage{cite}
\usepackage[colorlinks]{hyperref}
\hypersetup{pageanchor=false}
\usepackage{indentfirst}
\usepackage{url}
\usepackage{float}
\usepackage{caption}
\usepackage[numbers,square,comma,sort&compress,merge]{natbib}

\def\a{\alpha}

\def\d{\delta}
\def\D{\Delta}
\def\e{\varepsilon}

\def\f{\frac}

\def\l{\left}

\def\mc{\mathcal}

\def\m{\mu}
\def\n{\nu}

\def\nn{\nonumber}

\def\p{\partial}

\def\r{\right}
\def\s{\sigma}
\def\t{\theta}

\def\be{\begin{equation}}
\def\ee{\end{equation}}

\def\bea{\begin{eqnarray}}
\def\eea{\end{eqnarray}}

\def\ba{\begin{array}}
\def\ea{\end{array}}

\def\bc{\begin{center}}
\def\ec{\end{center}}

\def\bl{\begin{flushleft}}
\def\el{\end{flushleft}}

\def\br{\begin{flushright}}
\def\er{\end{flushright}}

\def\bi{\begin{itemize}}
\def\ei{\end{itemize}}

\def\bt{\begin{tabular}}
\def\et{\end{tabular}}

\newtheorem{question}{Question}
\def\bq{\begin{question}}
\def\eq{\end{question}}

\newtheorem{definition}{Def}
\def\bd{\begin{definition}}
\def\ed{\end{definition}}

\newtheorem{answer}{Answer}
\def\ban{\begin{answer}}
\def\ean{\end{answer}}

\newtheorem{possibleanswer}{Possible answer}
\def\bpa{\begin{possibleanswer}\normalfont}
\def\epa{\end{possibleanswer}}

\newtheorem{theorem}{Theorem}
\def\bth{\begin{theorem}}
\def\eth{\end{theorem}}

\begin{document}

\begin{titlepage}
\vspace{5cm}

\vspace{2cm}

\begin{center}
\bf \Large{Gauging nonrelativistic field theories  using the coset construction}

\end{center}

\begin{center}
{\textsc {Georgios K. Karananas, Alexander Monin}}
\end{center}

\begin{center}
{\it  Institute of Physics, \\
\'Ecole Polytechnique F\'ed\'erale de Lausanne, \\ 
CH-1015, Lausanne, Switzerland}
\end{center}

\begin{center}
\texttt{\small georgios.karananas@epfl.ch} \\
\texttt{\small alexander.monin@epfl.ch} 
\end{center}

\vspace{2cm}

\begin{abstract}

We discuss how nonrelativistic spacetime symmetries can be gauged in the context of the coset construction. We consider theories invariant under the centrally extended Galilei algebra as well as the Lifshitz one, and we investigate under what conditions they can be supplemented by scale transformations. We also clarify the role of torsion in these theories.

\end{abstract}

\end{titlepage}

\section{Introduction}

The necessary ingredients for building an effective field theory are the field/particle content and symmetries. The latter impose constraints on a Lagrangian, for it (or better to say the action) should be a singlet under the symmetry transformations. Once all the symmetries of a system are known, the number of free parameters in the Lagrangian is reduced.

The reason why it may be needed to go from rigid symmetries to gauged ones is twofold. On the one hand, the background gauge fields act like sources for the corresponding conserved currents. Gauge invariance in this case puts severe constraints (selection rules) on the partition function:  integrating out dynamical fields leads -- in the absence of anomalies -- to a gauge-invariant partition function. On the other hand, the gauge field theories are an appropriate language to talk about massless vector and tensor degrees of freedom, e.g. photons and gravitons.

Any global symmetry group can be made local by introducing a sufficient number of corresponding compensators (gauge fields) with appropriate transformation properties.\footnote{Strictly speaking, this is true only when the symmetry is not anomalous.} A question that naturally arises is whether this number can be smaller than the number of generators of the symmetry group considered. For internal symmetries (the ones that commute with the generators of spacetime translations), this does not seem to be the case. However, for spacetime symmetries the gauging may not require as many fields as there are generators. For example, the Poincar\'e group can be made local without introducing the spin connection as an independent field, but rather as a function of the vielbein (at least for torsionless theories). Also some Weyl invariant theories do not require the introduction of a gauge field to account for the local scale transformations, since its role can be played by a certain combination of curvature tensors~\cite{Iorio:1996ad,Karananas:2015eha}. 

In this paper, we focus on gauging nonrelativistic spacetime symmetries, namely the centrally extended Galilei algebra (also known as Bargmann algebra) and the Lifshitz algebra. nonrelativistic theories coupled to curved backgrounds appear naturally in Lorentz violating modifications of gravity, like Ho\v{r}ava-Lifshitz gravity~\cite{Horava:2009uw}, as well as in holographic duals of nonrelativistic systems~\cite{Balasubramanian:2008dm}. There has been renewed interest in these theories in the context of many body systems/condensed matter physics, which has been partially sparked by~\cite{Son:2005rv,Son:2008ye}. Even though there is a large number of papers dedicated to studying these systems~\cite{Son:2013rqa,*Geracie:2014nka,*Iorio:2014pwa,Andringa:2010it,Christensen:2013lma,Bergshoeff:2014uea,Hartong:2015zia,*Hartong:2015wxa,*Hartong:2014pma,*Hartong:2014oma,Brauner:2014jaa,Jensen:2014aia,Banerjee:2014pya,*Banerjee:2014nja,Geracie:2015xfa,*Geracie:2015dea}, we nevertheless believe that our approach allows one to clarify some subtleties.

The action of spacetime symmetries on the fields, obtained as an induced representation, is related to the nonlinear realization of symmetries. Therefore, when talking about these systems we find that the coset construction provides the appropriate language. It allows one to circumvent certain difficulties related to the transformation properties of the fields under the Galilei group, automatically providing the necessary building blocks. As a result, one gets a theory with local Galilei invariance without spontaneously breaking boosts and/or $U(1)$, as it was done for example in~\cite{Brauner:2014jaa,Hartong:2015wxa}. Moreover, within our approach, it is clear that for theories with local Galilei invariance the condition for vanishing spatial torsion is not consistent unless the temporal part of the torsion is set to zero as well.

One of our goals is to try to generalize the results of the paper~\cite{Iorio:1996ad} for the case of theories exhibiting local nonrelativistic invariance. Namely, we wish to understand the conditions under which a theory can be rendered Weyl invariant without introducing 
an additional gauge field $W_\m$ corresponding to local scale transformations.\footnote{We use greek letters $(\m,\n,\ldots)$ to denote spacetime indices.} With the coset construction, it is straightforward to show~\cite{Karananas:2015eha} that if for Lorentz invariant theories the field $W _ \m$ appears only in a very specific combination, it can be traded for Ricci curvatures (in this case it is said that the theory can be \emph{Ricci gauged}). 

Using the same approach, we address this question for the case of nonrelativistic theories coupled to a curved background. Considering first the centrally extended Galilei algebra, we show that the mere notion of Weyl invariance can be introduced only for torsionful theories. We show that for twistless torsionful theories, it is always possible to express the spatial components of the Weyl vector in terms of torsion, which in turn is a function of the vielbein.

Next, we turn to the Lifshitz algebra. In this case, there is no obstacle to the complete elimination of the Weyl gauge field; thus, any scale invariant theory in flat space can be coupled to a curved background in a Weyl invariant way, provided one allows for nonvanishing torsion. This is  similar to the situation occurring with Lorentz invariant theories~\cite{Karananas:2015eha},  where torsion may play the role of an additional degree of freedom making a theory Weyl invariant.

This article is organized as follows. In Sec.~\ref{sec:coset_constr}, we show how spacetime symmetries can be gauged within the coset construction. In Sec.~\ref{sec:Galilei}, we gauge the Galilei algebra and we demonstrate how matter fields can be coupled systematically to curved backgrounds. Moreover, we show what the constraints leading to torsionless and torsionful geometries are. In Sec.~\ref{sec:gal}, we study the scale invariant generalizations of the Galilei as well as the Lifshitz algebras. For the former, by solving the inverse Higgs constraint, we express the spatial part of the vector field associated with scale transformations in terms of the vielbein. In addition, we demonstrate that locally Lifshitz-invariant theories can always be made Weyl invariant without introducing the corresponding independent gauge field. We present our conclusions in Sec.~\ref{sec:conclusions}.

\section{Gauging spacetime symmetries}
\label{sec:coset_constr}

In this section, we discuss the relevance of the coset construction for gauging spacetime symmetries. The nonlinear realization of internal symmetries was introduced in~\cite{Coleman:1969sm,*Callan:1969sn} and it is used to obtain the building blocks for a theory that exhibits a specific symmetry breaking pattern $S \to S _ 0$. In other words, it allows one to construct the most general action of a group $S$ such that when restricted to its subgroup $S _ 0$, it becomes a linear representation. 

The procedure can be briefly described as follows. For the symmetry breaking pattern, one realizes the action of the group $S$ on the coset space $S/S _ 0$ by left multiplication. Choosing the coset representative as 
\be
\Omega = e ^ {i \pi T} \in S \ ,
\label{coset_rep}
\ee
where $T$ is the set of all broken generators and $\pi$ (Goldstone fields) constitutes a parametrization of the coset,\footnote{For brevity we suppress all the indices corresponding to the Lie algebra.} one gets the transformation
\be
s \Omega = \Omega ' \bar s _ 0\ , ~~~\text{with}~~~ \bar s_0\equiv \bar s _ 0 (\pi,s) \in S _ 0 \ .
\label{g_action_coset}
\ee 
If we denote by $t$ all the unbroken generators, it is easy to check that the Maurer-Cartan form
\be
\Omega ^ {-1} \p _ \m \Omega = i \nabla _ \m \pi \, T + i \omega _ \m t \ ,
\label{MC_form}
\ee
transforms under~\eqref{g_action_coset} as
\be
\l ( \Omega ^ {-1} \p _ \m \Omega\r ) ' = \bar s _ 0 \l( \Omega ^ {-1} \p _ \m \Omega \r ) \bar s _ 0 ^ {-1}  + \bar s _ 0 \p _ \m \bar s _ 0 ^ {-1} \ .
\ee
For compact groups, the above translates into the corresponding transformations of  $\nabla_\m \pi$ and $\omega_\m$
\be
\begin{aligned}
\nabla_\m \pi ' T & =  \bar s _ 0  \, \nabla_\m \pi ' T \, \bar s _ 0 ^ {-1}  \ ,  \\
i\omega _ \m ' t & =  \bar s _ 0  \, i\omega _ \m  t \, \bar s _ 0 ^ {-1} + \bar s _ 0 \p _ \m \bar s _ 0 ^ {-1} \ ,
\label{h_trans_coset}
\end{aligned}
\ee
which can be used to write automatically $S$-invariant Lagrangians by constructing singlets of the subgroup $S _ 0$. The gauging of the group 
$S$ is achieved by generalizing the partial derivative in~\eqref{MC_form} to a covariant derivative including gauge fields that under the action of 
$S$ transform as 
\be
\tilde A ' _ \m = s \tilde A _ \m s ^ {-1} + s \p _ \m s ^ {-1} \ .
\label{gauge_fields_trans}
\ee

The difference between internal and spacetime symmetries is that the latter are usually (if not necessarily) realized on the infinite dimensional spaces of fields. These infinite dimensional representations are induced representations that are defined in the following way. For a group $K$, its subgroup $K _ 0 \subset K$ that is realized on a linear space $V$, there is a natural action of the group $K$ on the coset $K  / K _ 0$ by left multiplications.\footnote{Usually the coset $K  / K _ 0$ is isomorphic to the spacetime manifold.} Viewing the group $K$ as a fiber bundle with base $K / K _ 0$, one realizes its action on the space of sections of the associated bundle with fibers isomorphic to $V$. For example, let us take $K$ to be the $n$-dimensional Poincar\'e group and $K _ 0$ to be the Lorentz group. It is clear that in this case $K / K _ 0 = \mathbb{R} ^ n$. The action of $K$ on the coset is as follows
\be
k e ^ {i P y} = e ^ {i P (\Lambda y + a)} \bar k _ 0 (k) \ ,
\ee
where $k \in K$, $\bar k _ 0 (k) \in K _ 0$, $P_A$ are momenta, $\Lambda_A$ correspond to Lorentz rotations, $y_A$ are Cartesian coordinates on the coset~$\mathbb{R} ^ n$ and $a_A$ are parameters of the translations. Considering a representation of the Lorentz group
\be
\begin{aligned} 
\rho: K _ 0 & \to & {GL} ( V ) \ ,  \\
T _ {k _ 0} \Psi & = & \rho (k_0) \Psi \ ,
\end{aligned}
\ee
we define the induced representation of the full Poincar\'e group according to
\be
\l ( T _ k \Psi \r ) (y') = \rho (\bar k _ 0 (k)) \Psi (y) \ ,
\ee
which corresponds to the standard transformation of a field 
\be
\Psi _ {(\Lambda,a)} (y) = D (\Lambda) \Psi (\Lambda ^ {-1} y - a) \ .
\ee

Even though the generators $P _ A$, which correspond to the coset $K / K _ 0$, are not  broken and are realized linearly on the space of fields, the very construction of this representation makes it natural to include the momenta in the coset~\eqref{coset_rep} when discussing the breaking and/or gauging of spacetime symmetries. Consequently, for the symmetry group $G$ that includes both internal and spacetime symmetries and that is broken down to a subgroup $H$, one gets the coset in the form
\be
\Omega = e ^ {i P x} e ^ {i \pi (x) T} \ ,
\ee
where by $T$ we denote all the broken generators (not only the internal ones), whereas by $t$ we represent all the unbroken ones apart from the momenta $P$. The way to introduce a different set of coordinates on the spacetime manifold  is to have them appearing in the coset representative through the functions $y^A (x)$, which means that in general we may write
\be
\Omega = e ^ {i P y (x)} e ^ {i \pi (x) T} \ .
\ee
Under the action of the spacetime symmetry group $K$, the coordinates transform according to
\be
k e ^ {i P y (x)} = e ^ {i{P y (x')}} \bar k _ 0 \ ,~~~\text{with}~~~ \bar k _ 0\equiv \bar k _ 0(x,k) \ .
\ee
These transformations may be viewed in a different way, namely, keeping the coordinates $x$ unchanged while transforming the functions $y^A (x) \to y^{'A} (x)$ \footnote{For example, in a two-dimensional Euclidean space, one may choose polar coordinates corresponding to $(y^1 (r,\varphi), y^2(r,\varphi)) = (r \sin \varphi, r \cos \varphi)$. Then the transformation under rotations 
\be
(y^1, y^2) \to (r \cos (\varphi + \a), r \sin (\varphi + \a)) \ , \nn
\ee
can be equivalently viewed either like
$\varphi \to \varphi' = \varphi + \a$, or as a change of the functional form $y ^ {'1} (r,\varphi)= r \cos (\varphi + \a)$, and similar for $y ^ {'2}$.}
\be
k e ^ {i P y (x)} = e ^ {iP y' (x)} \bar k _ 0 \ .
\ee
The reason for this choice becomes clear when the gauging of a spacetime symmetry group is considered, for in this case one does not have to take into account the transformation of the fields due to the change of coordinates $x$ and the gauging goes along the lines of that for internal symmetries. However, by doing so, the additional functions $y^A(x)$, with very specific transformation properties, had to be introduced. Of course, they are not physical and should be dispensed with. This is easily achieved by simply demanding that the resulting theory is invariant under diffeomorphisms as well. 

In a sense, introducing these additional spurious fields allows us to decouple the diffeomorphisms from the (local) transformations under the spacetime symmetry group. The gauge fields $ \tilde A _ \m$ transform in the standard way~\eqref{gauge_fields_trans} under the local spacetime transformations and separately under the diffeomorphisms $x \to x'$,
\be
\tilde A '_ \m (x') = \tilde A _ \n (x) \f {\p x ^ \n} {\p x ^{' \m}} \ . 
\label{gauge_diffeomorphisms}
\ee
The Maurer-Cartan form can now be written as
\be
\Omega ^ {-1} \tilde D _ \m \Omega 
 =  i  e_\m^A P_A+ i \nabla _ \m \pi \, T + i \omega _ \m t \ ,
\ee
where as before $P _ A$ are momenta, whereas $t$ and $T$ are the rest of the unbroken and broken generators respectively. For symmetry groups with the following schematic structure of commutation relations
\be
\begin{aligned}
\l [ t,t \r ] & = t \ , \\
\l [ t, P \r ] & =  P \ ,  \\
\l [ t, T \r ] & =  T \ , 
\label{representation_commut}
\end{aligned}
\ee
and upon using the definition of the transformation of the coset representative
\be
g \Omega = \Omega' \bar h (y,g) \ ,
\label{coset_rep_trans}
\ee
we find that the transformations of $\nabla _ \m \pi$, $\omega _ \m$ and $e_ \m ^ A$, are given by
\be
\begin{aligned}
\nabla \pi ' T & =  \bar h \, \nabla \pi ' T \, \bar h (\pi,s) \ , \\
i\omega _ \m ' t & =  \bar h \, i \omega _ \m  t \, \bar h ^ {-1}+ \bar h \p _ \m \bar h ^ {-1} \ ,  \\
e ^ {'A} _ \m P _ A & =  e ^ {A} _ \m \, \bar h (\pi,g) P _ A \bar h ^ {-1} (\pi,g) \ ,
\label{space_time_h_trans_coset}
\end{aligned}
\ee
The coefficients $e_ \m ^ A$ due to their specific transformation properties under the diffeomorphisms~\eqref{gauge_diffeomorphisms} can be thought of as the vielbein.

\section{Galilei algebra}
\label{sec:Galilei}

The coset construction techniques have already been used to gauge the Poincar\'e group in~\cite{Ivanov:1981wn,*Ivanov:1981wm,*Delacretaz:2014oxa}, as well as the Galilei group in~\cite{Brauner:2014jaa}, where Goldstone bosons for boosts were introduced. Here we consider the gauging of the Galilei group but without spontaneously breaking any symmetry. This possibility was mentioned in~\cite{Brauner:2014jaa} and partly worked out in \cite{Jensen:2014aia}.

The centrally extended Galilei algebra (sometimes called Bargmann algebra) in a $n$-dimensional spacetime can be obtained from the Poincar\'e one using the standard \.In\"on\"u-Wigner contraction~\cite{Inonu:1953sp}. The nonvanishing commutation relations are given by
\be
\begin{aligned}
&\l [ J _ {i j}, J _ {k l} \r ]  =  i \l ( J _ {j l} \d _ {i k} + J _ {i k} \d _ {j l} - J _ {i l} \d _ {j k} - J _ {j k} \d _ {i l}  \r ) \ ,  \\
&\l [ J _ {i j}, P _ {k} \r ] = i \l ( \d _ {i k} P _ j - \d _ {j k} P _ i  \r ) \ ,  \\
&\l [ J _ {i j}, K _ {k} \r ]  =  i \l ( \d _ {i k} K _ j - \d _ {j k} K _ i  \r ) \ ,  \\
&\l [ K _ {i}, P _ {j} \r ] = - i \d _ {i j} M \ ,  \\
&\l [ K _ {i}, H \r ]  =  - i P _ i \ .
\label{Galilei_cr}
\end{aligned}
\ee
In the above, $J$ correspond to (spatial) rotations, $K$ correspond to boosts, $H$ and $P$ correspond to temporal and spatial translations respectively, and $M$ is the central extension corresponding to the particle number operator or the mass.

To build a theory with local Galilei invariance, we consider the coset space of the full Galilei group $\mathrm{Gal} (n)$ over its subgroup generated by $J$, $K$ and $M$. Following the logic described in the previous section, we take the coset representative in the form
\be
\Omega = e ^ {i H z + i P _ i y ^ i}\ .
\label{coset_momenta}
\ee
Introducing the gauge fields $\tilde n_\m$ and $\tilde e_\m^i$ for temporal and spatial translations, respectively, $\tilde \omega^i_\m$ for boosts, $\tilde \t ^ {i j} _ \m$ for $SO(n-1)$ rotations, and $\tilde A_\m$ for the particle number $U (1)$, we find that the Maurer-Cartan form is given by the following expression,
\be
\Omega ^ {-1} \tilde D _ \m \Omega 
 =  i n _ \m H +i e_\m^i P_i+ i \omega ^ i _ \m K _ i+ \f {i} {2} \t ^ {i j} _ \m J _ {i j} + i A _ \m M \ ,
\ee
where the quantities without the tilde could be thought of as the fields in the unitary gauge. According to the procedure described in the previous section, the fields $n _ \m$ and $e _ \m ^ i$ are identified with the temporal and spatial components of the vielbein. For later convenience, we also define the inverse vielbein,\footnote{The existence of the inverse vielbein is guaranteed by the fact that $\det \l ( \p _ \m y ^ A \r ) \neq 0$.} $ V^\m\equiv E ^ \m _ 0$ and $E^\m_i$, such that
\be
V^\m n_\m=1 \ ,~~~V^\m e_\m^i =0 \ ,~~~n_\m E^\m_i=0 \ ,~~~e_{\m i} E^\m_j=\d_{ij} \ ,~~~e_\m^i E^\n_i=\d^\n_\m-n_\m V^\n \ .
\ee
The transformation properties of the fields can be obtained from the transformation of the coset representative~\eqref{coset_rep_trans}. However, the structure of the commutation relations of the Galilei group~\eqref{Galilei_cr} is not the one presented in~\eqref{representation_commut}. This fact results in the mixing of the $U (1)$ gauge field with the vielbein under boosts. In the following table, we present the transformation properties of the fields under rotations $J$, boosts $K$ and $U(1)$ with parameters $ R _ {i j}$, $\eta_i$, and $\a$ correspondingly.
\begin{table}[H]
\centering
\bt{c | cccc}
 &$J$  & $K$ & $M$\\
\hline
$n ' _ {\m}$ & $n _ {\m}$   &  $n _ \m$ & $n _ {\m}$ \\
$V ^ {' \m}$ & $V^ {\m}$   &  $V^\m+ \eta^ i E ^ \m _ i$ & $V _ {\m}$ \\
$e ^ {'i} _ {\m}$ & $R_{ij } e^ {j} _ {\m}$ & $e _ \m ^ i - \eta ^ i n _ \m$ & $e _ \m ^ i$ \\
$E^{'\m}_i$ & $R _ {i j} E^ {\m} _j$ & $E^\m _i$ & $E ^ \m _ i$ \\
$\t ^ {'i j} _ {\m}$ & $R _ {i k} R_ {j l} \t ^ {kl} _ {\m} + \l ( R \p _ \m R ^ {-1}\r ) _ {i j}$ & $\t ^ {i j} _ {\m}$ & $\t ^ {i j} _ {\m}$ \\
$\omega ^ {'i} _ {\m}$ & $R _ {i j} \omega ^ {j} _ {\m}$ & $\omega ^ {i} _ {\m} + \t ^ {i j} _ {\m} \eta _ j + \p _ \m \eta _ i$ & $\omega ^ {i} _ {\m}$ \\
$A _ \m '$ & $A _ \m $ & $A _ \m - \eta _ i e ^ i _ \m + \f {1} {2} \eta^2 n _ \m$ & $A _ \m + \p _ \m \a $
\et
\label{gfield_trans_1}
\end{table}
It should be noted that the actual transformation properties of $A_\m$ are different from the ones presented in  the above table.
Indeed, using the commutation relations of the Galilei group, it is straightforward to show that
\be
e ^ {-i K \eta} e ^ {i P y} = e ^ {i P y'} e ^ {-i K \eta} e ^ {-i M f} \ ,~~~\text{ with} ~~~f = \eta _ i y ^ i  - \f {1} {2} \eta ^ 2 z  \ .
\ee
Hence, the ``honest'' transformation of the $U (1)$ gauge field under $K$ is given by
\be
A ' _ \m = A _ \m - \eta _ i e ^ i _ \m + \f {1} {2} \eta^2 n _ \m + \p _ \m f \ .
\ee
The last term in the expression above was dropped in the previous table, since  it has precisely the form of the gauge transformation of $A_\m$.  

The standard definition of the field strengths leads to
\be
\begin{aligned}
\label{field_str}
 n _ {\m \n} & = 2 \p _ {[\m}n _ {\n]} \ , \\
 e ^ i _ {\m \n} & =  2\p _ {[\m}e ^ i _ {\n]} + 2\t _ {[\m} ^ {i j} e _ {\n]  j} 
+2\omega _ {[\m} ^ {i} n _{ \n]} \ , \\
\t ^ {i j} _ {\m \n} & =  2\p _ {[\m} \t ^ {i j} _ {\n]} + \t ^i_ {\m k} \t^{kj} _ \n-  \t ^i_ {\n k} \t^{kj} _ \m \ , \\
\omega ^ {i} _ {\m \n} & =  2\p _ {[\m} \omega ^ {i} _ {\n]}
+ 2\t _ {[\m} ^ {i j} \omega _ {\n] j} \ , \\
A _ {\m \n} & =  2\p _ {[\m} A _{ \n]} + 2\omega _ {[\m} ^ {i} e   _ {\n] i}\ ,
\end{aligned}
\ee
where the brackets $[\ldots]$ denote antisymmetrization of the corresponding indices. A straightforward calculation reveals that
\begin{table}[H]
\centering
\bt{c | cccc}
 &$J$ & $K$ & $M$\\
\hline
$n' _ {\m \n}$ & $n _ {\m \n}$   &  $n _ {\m  \n}$ &  $n _ {\m  \n}$ \\
$e ^ {'i} _ {\m  \n}$ & $R _ {i j} e^ {j} _ {\m  \n}$ & $e _ {\m  \n} ^ i - \eta ^ i n _ {\m  \n}$ &  $e _ {\m  \n} ^ i$ \\
$\t ^ {'i j} _ {\m  \n}$ & $R _ {i k} R _ {j l} \t ^ {kl} _ {\m  \n} $ & $\t ^ {i j} _ {\m  \n}$ & $\t ^ {i j} _ {\m  \n}$ \\
$\omega ^ {'i} _ {\m  \n}$ & $R _ {i j} \omega ^ {j} _ {\m  \n}$ & $\omega ^ {i} _ {\m  \n} + \t ^ {i j} _ {\m  \n} \eta _ j $ &  $\omega ^ {i} _ {\m  \n}$ \\
$A _ {\m  \n} '$ & $A _ {\m  \n} $ & $A _ {\m  \n} - \eta _ i e ^ i _ {\m  \n} + \f {1} {2} \eta _ i \eta _ i n _ {\m \n}$ & $A _ {\m  \n}$
\et
\label{gfield_trans_2}
\end{table}

\subsection{Coupling to matter}

With the gauge fields at our disposal, we can build the temporal and spatial \emph{covariant derivatives} of a matter field $\Psi$ belonging to an irreducible\,\footnote{These are induced by representations of the $SO(n-1)$ rotation group. In our case the action of boosts on matter fields is trivial.} representation of the Galilei group as
\be
\begin{aligned}
\nabla _ t \Psi &= V^ \m \l ( \p _ \m \Psi + \f {i} {2} \t ^ {i j} _ \m \rho (J _ {i j}) \Psi + i m A _ \m \Psi \r ) \ ,\\
\nabla _ i \Psi &= E ^ \m _ i \l ( \p _ \m \Psi + \f {i} {2} \t ^ {j k} _ \m \rho (J _ {j k}) \Psi + i m A _ \m \Psi \r ) \ ,
\label{covariant_D}
\end{aligned}
\ee
where  $\rho$ is the representation of the $\mathfrak{so}(n-1)$ the field belongs to, and $m$ is the charge of  $\Psi$ under $U(1)$. It can be easily shown that at the leading order in $\eta$
\be
\begin{aligned}
( \nabla _ t \Psi ) ' & =  \nabla _ t \Psi + \eta _ i \nabla _ i \Psi \ , \\
( \nabla _ i \Psi ) ' & =  \nabla _ i \Psi - i m \, \eta _ i \nabla _ t \Psi\ .
\label{matter_boosts}
\end{aligned}
\ee
Even though the derivatives defined above do not actually transform covariantly under local boosts it is still true that any Lagrangian that is invariant under the Galilei group in flat space -- which corresponds to the limit where all gauge fields vanish -- can be made locally Galilei invariant by substituting all partial derivatives by covariant ones, i.e. $\p _ t \to \nabla _ t$ and $\p_i\to\nabla_i$. The local invariance of the Lagrangian under rotations and $U (1)$ is clear, for under their action, the covariant derivative transforms covariantly. The only nontrivial point is the transformation with respect to boosts, which in the flat background has the form
\be
\Psi ' (t,x)= e ^ {-i m \, x ^ i v ^ i} \Psi (t,x _ i + v _ i t) \ , ~~~\text{with}~~~ v _ i = const \ . \\
\ee
Let us consider a Lagrangian that is invariant under boosts, i.e.
\be
\mc  L \l [ \p _ t \Psi, \p _ i \Psi, \Psi \r] =
\mc  L \l [e ^ {-i m \, x ^ i v ^ i} \l ( \p _ t \Psi + v _ i \p _ i \Psi \r ),e ^ {-i m \, x ^ i v ^ i} \l ( \p _ i \Psi - i m \, v _ i \p _ t \Psi \r ), \Psi (t,x) \r] \ ,
\label{Lagr_flat_boost}
\ee

It follows automatically that the Lagrangian with all partial derivatives substituted by covariant ones is invariant under local boosts. Indeed, the fact that the transformations~\eqref{matter_boosts} coincide in the flat limit [up to the $U (1)$ factor that we dropped] with the ones presented implicitly in~\eqref{Lagr_flat_boost} guarantees the cancellation of all factors containing $\eta$ (there are no terms that contain derivatives of $\eta$).

For example, consider the theory of a field $\psi$ with spin $s$ in a $2+1$-dimensional flat spacetime whose dynamics is described by the following Lagrangian 
\be
\mc L = \f {i} {2} \bar \psi \overset{\leftrightarrow}{\p} _ t \psi - \f {1} {2 m} \partial _ i \bar \psi \partial _ i  \psi \ ,
\label{flat_free_Lagr}
\ee
with $\bar \psi \overset{\leftrightarrow}{\p} _ t \psi = \bar \psi \p _ t  \psi - \p _ t \bar \psi \psi$. Promoting partial derivatives to covariant ones and multiplying by the determinant of the temporal and spatial vielbeins (denoted collectively by $\det e$), we obtain the action that is locally Galilei and diffeomorphism invariant,
\be
S = \int d t d^2 x \,  \det e \, \l ( \f {i} {2} \bar \psi \overset{\leftrightarrow}{\nabla} _ t \psi - \f {1} {2 m} \nabla _ i \bar \psi \nabla _ i  \psi \r ) \ .
\label{free_curved}
\ee
It should be stressed that  had we chosen a Lagrangian with the time derivative appearing in the nonsymmetric form, i.e.
\be
\mc L _ {\text{nonsym}}= i \bar \psi \p _ t  \psi - \f {1} {2 m} \partial _ i \bar \psi \partial _ i  \psi \ ,
\ee
which differs from~\eqref{flat_free_Lagr} by $-\f {i} {2} \p _ t \l ( \bar \psi \psi \r )$, the procedure would not have worked. The reason is that the Lagrangian in this case is not invariant under boosts, but rather it shifts by a total derivative. From~\eqref{Lagr_flat_boost}, it follows that $\D \mc L _ {\text{nonsym}} = -\f {i} {2} v _ i \p _ i \l ( \bar \psi \psi \r )$, which  cannot be written as a total derivative upon promoting $v _ i$ to $\eta _ i (x)$, since $\p \eta$ terms do appear in this case.

\subsection{Torsionless geometry}

At the moment, we have all the building blocks for constructing a theory with local Galilei symmetry. However, it appears that there are many more degrees of freedom than are actually needed in order to accomplish our goal. The standard way to eliminate redundancies within the coset construction is to impose covariant constraints that can be solved algebraically. 

Using the transformation properties of the fields, we see that the only covariant quantity is the temporal component of the torsion $n _ {\m \n}$. Meanwhile, both the spatial torsion $e ^ i _ {\m \n}$ and the $U (1)$  field strength $A _ {\m \n}$ transform covariantly under all group operations, apart from boosts. However, the mixing of $e ^ i _ {\m \n}$ and $A _ {\m \n}$ with $n _ {\m \n}$ can be eliminated by imposing
\be
n _ {\m \n} = 0 \ .
\label{temporal_torsion}
\ee
It is clear that in this case $n_\m$ corresponds to a closed form, i.e. $n _ \m = \p _ \m \tau$ where $\tau$ is some function that can be identified with global time. With this condition, the other two constraints,
\be
e ^ i _ {\m \n} = 0 
\label{spatial_torsion}
\ee
and
\be
A _ {\m \n} = 0 \ ,
\ee
become covariant and they can be used to specify completely the  $\mathfrak{so}(n-1)$ part of the connection
\be
\label{thet_2}
\t^{ij}_\m = \bar \t^{ij}_\m \equiv \p _ {\l [ \m \r.} e _ {\l. \n \r ]} ^ j E ^ {\n} _ i -\p _ {\l [ \m \r.} e _ {\l. \n \r ]} ^ i  E ^ {\n } _ j - \p _ {\l [ \rho \r.} e _ {\l. \s \r ]} ^ k e_{\m k} E ^{\rho } _ iE ^{ \s } _ j +n_\m E ^ {\rho} _ i E ^{ \s} _ j \p _ {\l [ \rho \r.} A _ {\l. \s \r ]}-2e_{\m[i}E^\rho_{j]}V^\s\p_{[\rho}n_{\s]} \ ,
\ee
as well as the connection that corresponds to boosts
\be
\label{omeg_2}
\omega ^ i _ \m = \bar \omega ^ i _ \m \equiv 2E ^ {\s} _ i V ^ \n \p _ {\l [ \s \r.} A _ {\l. \n \r ]} n _ \m + \l ( E ^ {\s} _ i E ^ {\n} _ j \p _ {\l [ \s \r.} A _ {\l. \n \r ]} + 2 \p _ {\l [ \s \r.} e _ {\l. \n \r ]} ^ {\l ( i\r.} E ^ {\l. j \r ) \s} V ^ \n \r ) e ^ j _ \m \ ,
\ee
where the parentheses $( \ldots )$ denote symmetrization.

Continuing with the example that we started previously, we see that the term corresponding to the interaction of the spin and the magnetic field  appears naturally in the action~\eqref{free_curved}. Indeed, using the expression~\eqref{thet_2}, we see from the first term in~\eqref{free_curved} that the derivative of the gauge field $A _ \m$ couples to $\bar \psi \psi $
as
\be
 \f {i} {2} \bar \psi \overset{\leftrightarrow}{\nabla} _ t \psi \supset - \f {s} {2} \e _ {i j}  E ^ {\m} _ i E ^ {\n} _ j \p _ {\l [ \m \r.} A _ {\l. \n \r ]}\bar\psi \psi \ .
\ee
Upon an appropriate rescaling of the fields, the coupling constant $g _s$ appears in front of this term. There is no need for a redefinition of the transformation properties of the gauge field $A _ \m$ in order to make the theory invariant under the general coordinate transformations, as was done for example in~\cite{Geracie:2014nka,Jensen:2014aia}. A somewhat similar approach was suggested in~\cite{Andreev:2013qsa}.

\subsection{Torsionful theory}

It should be stressed that it is not consistent to impose the spatial torsionlessness condition~\eqref{spatial_torsion} without having the temporal torsion be zero as well, for the condition $e ^ i _ {\m \n} = 0$ alone is not invariant under boosts. However, there is still an alternative to what was done in the previous section. According to the coset construction, any covariant constraint can be imposed without contradicting the symmetry breaking pattern. The tensor $n _ {\m \n}$ can be naturally decomposed into representations of the $\mathfrak{so}(n-1)$, namely, 
$E ^ \m _ i E ^ \n _ j n _ {\m \n}$ and $E ^ \m _ i V ^ \n n _ {\m \n}$. However, only the first one is a singlet with respect to the boosts and thus can be safely set to zero,
\be
E ^ \m _ i E ^ \n _ j n _ {\m \n} = 0 \ .
\label{ttnc_n}
\ee
The constraints consistent with the above condition are the following:
\be
e ^ i _ {\m \n} E^{\m}_i E^\n_j = 0 ~~~ \text {and} ~~~ E^{\m}_i E^\n_j A_{\m\n} = 0 \ .
\ee
Consequently, the spin connection $\t ^ {ij} _ \m$ and $\omega ^ i _ \m$ can be fixed only partly, since we can express in terms of the vielbein and the U(1) gauge field only $(n-1)^2(n-2)/2 + (n-2)(n-1)/2$ components. These  correspond to $\t^{ij}_\m E ^ \m _ k$ and $\omega^ {\l [i \r.} _ \m E ^ {\l. j \r ] \m}$ respectively.

We should also note that the condition~\eqref{ttnc_n} coincides with the one imposed on the temporal torsion in the case of the twistless torsional Newton-Cartan (TTNC) geometry discussed in a number of papers~\cite{Christensen:2013lma,Hartong:2014oma,Hartong:2014pma,Bergshoeff:2014uea,Hartong:2015wxa,Hartong:2015zia}. Contrary to our case, the authors of~\cite{Hartong:2015wxa,Hartong:2014oma} were able to fully determine the connections associated with spatial rotations and boosts. This was made possible by introducing a ``St\"uckelberg field," thus requiring that the $U(1)$ symmetry be realized nonlinearly.

\section{Adding dilatations}
\label{sec:gal}

\subsection{Galilei algebra}

It is interesting to investigate under what conditions a theory that is scale invariant in flat space can be promoted to a Weyl invariant one without   introducing a gauge field corresponding to the local scale transformations. Notice that the nonzero commutators of the dilatation generator $D$ and the Galilei ones are 
\be
\begin{aligned}
&\l [ D, H \r ]  = - 2 i H \ , \l [ D, P _ i \r ]  =  - i P _ i \ , \l [ D, K _ i \r ]  =  i K _ i \ .
\end{aligned}
\ee
As one can see, the scaling of space and time for theories that are not Lorentz invariant does not have to be homogeneous, which is manifest due to the factor $2$.

At this point we have to decide what geometry to consider. It is rather obvious that the standard transformation
\be
n_ \m \to e ^ {- 2 \s} n _ \m \ ,
\ee
is not consistent with the torsionlessness condition $\p _ {[\m}n _ {\n]} = 0$.  The other option is~\eqref{ttnc_n}, which as we saw leads to additional -- as compared to the Newton-Cartan data~\cite{Son:2013rqa} -- independent degrees of freedom.

The coset construction provides the natural language to speak about  local scale transformations as well. The only modification one has to make to the procedure used for gauging the Galilei algebra is to introduce yet another gauge field $W _ \m$ that corresponds to the dilatations. The transformation properties of the fields under the Galilei group are not changed and are given in the tables of the previous section. The scaling properties may be found using the commutation relations presented previously. The ones that are not singlets are as follows:
\be
\hat n _ {\m} =  e ^ {2 \s} n_ \m \, , ~\, \hat V ^ {\m} = e ^ {- 2 \s} V^ \m \, , ~\, \hat e ^ {i} _ {\m} = e ^ {\s} e ^ {i} _ {\m} \, , ~\, \hat E^{\m}_i = e ^ {\s} E ^ \m _i \, , ~\, \hat \omega ^ {i} _ {\m} = e ^ {-\s} \omega ^ {i} _ {\m} \, , ~\, \hat W _ \m = W _ \m - \p _ \m \s \, .
\label{scaling_gauge}
\ee
Similarly, for the (modified) field strengths
\be
\begin{aligned}
\label{mod_field_str}
n _ {\m \n} & = 2 \p _ {[\m}n _ {\n]}+ 4W _ {[\m} n _ {\n]} \ , \\
e ^ i _ {\m \n} & =  2\p _ {[\m}e ^ i _ {\n]} + 2\t _ {[\m} ^ {i j} e _ {\n]  j} 
+2\omega _ {[\m} ^ {i} n _{ \n]}  +2 W _{[ \m} e ^ i _ {\n]} \ , \\
\t ^ {i j} _ {\m \n} & =  2\p _ {[\m} \t ^ {i j} _ {\n]} + \t ^i_ {\m k} \t^{kj} _ \n-  \t ^i_ {\n k} \t^{kj} _ \m \ , \\
\omega ^ {i} _ {\m \n} & =  2\p _ {[\m} \omega ^ {i} _ {\n]}
+ 2\t _ {[\m} ^ {i j} \omega _ {\n] j} - 2W _ {[\m} \omega ^ i _ {\n]} \ , \\
W _ {\m \n} & = 2 \p _ {[\m }W _ {\n]} \  ,  \\
A _ {\m \n} & =  2\p _ {[\m} A _{ \n]} + 2\omega _ {[\m} ^ {i} e   _ {\n] i}\ ,
\end{aligned}
\ee
we find that  
\be
\hat n  _ {\m \n} = e ^ {2 \s} n _ {\m \n}\ , ~~~ \hat e ^ {i} _ {\m \n} = e ^ {\s} e ^ {i} _ {\m \n}\ , ~~~ \hat \omega ^ {i} _ {\m \n} = \hat e ^ {-\s} \omega ^ {i} _ {\m \n} \ .
\label{scaling_strength}
\ee
Imposing the constraint~\eqref{temporal_torsion} does not lead to the torsionless geometry, i.e. the 1-form $n_\m$ is not forced to be closed, but rather it satisfies the TTNC condition~\eqref{ttnc_n}, which is compatible with the scaling transformations~\eqref{scaling_gauge}. On top of that, the constraint on the temporal torsion allows us to express the spatial part of the Weyl gauge field $W _ \m$ in terms of the vielbein. We readily obtain
\be
W _ i \equiv W _ \m E ^ \m _ i =-E ^ \m _ i V ^ \n \p _ {\l [ \m \r.} n _ {\l. \n \r ]} \ .
\label{spatial_W}
\ee

Solving the other two constraints~\eqref{spatial_torsion} in this case produces
\begin{align}
\label{spatial_theta_W}
\t^{ij}_\m & =  \bar \t^{ij}_\m + 2 e _ \m ^ {\l [ i \r.} W ^ {\l. j \r]} \ ,  \\
\omega ^ i _ \m & =  \bar \omega ^ i _ \m + W _ t e ^ i _ \m \ ,
\end{align}
where $\bar \t^{ij}_\m$ and $\bar \omega ^ i _ \m$ are given respectively by~\eqref{thet_2} and~\eqref{omeg_2}, and we defined the temporal component of the Weyl gauge field as $W _ t = V ^ \m W _ \m$. Having no other covariant quantities that we can use in order to eliminate $W _ t$, we can conclude that for generic curvature $\t ^ {i j} _ {\m \n}$, it is impossible to express the temporal part of the Weyl field in terms of the vielbein and $A _ \m$, so it stays an independent degree of freedom. However, this does not necessarily mean that a theory cannot be made Weyl invariant without introducing this additional degree of freedom.

Indeed, as before [see~\eqref{covariant_D}], the covariant derivative can be defined as
\be
D _ t \Psi = \nabla _ t \Psi - \D W _ t \Psi, ~~ D _ i \Psi = \nabla _ i \Psi - \D W _ i \Psi \ ,
\ee
where $\D$ is the scaling dimension of the field $\Psi$.\footnote{The Weyl transformation of a field has the form $\hat \Psi = e ^ {-\D \s} \Psi$.}  We see that if it is possible to rewrite the Lagrangian of a theory in flat spacetime such that the time derivative appears only in the ``symmetric way'' 
$\bar \psi \overset{\leftrightarrow}{\p} _ t \psi$, then in curved space this leads to 
\be
\bar \psi \overset{\leftrightarrow}{D} _ t \psi = \bar \psi \overset{\leftrightarrow}{\nabla} _ t \psi \ ,
\ee
which is independent of $W _ t$. As a result, such a theory is going to be automatically Weyl invariant, for the time derivative is the only source of
$W _ t$.

It is interesting to note that, as in relativistic theories, the presence of Weyl symmetry guarantees that when the flat spacetime limit is considered, the resulting theory is conformal.  The opposite, however, is not true (see~\cite{Karananas:2015ioa}). In the context of  Galilei-invariant theories, the conformal transformations are defined analogously to the relativistic case as the diffeomorphisms preserving the vielbein up to a conformal factor $\Omega$,
\be
\label{nr-conf-viel}
n'_\m=\Omega^2\,n_\m \ ,~~~e^{'i}_\m = \Omega\l(\Lambda^i_je^i_\m +\Lambda^i n_\m \r) \ ,
\ee
where $\Lambda^i_j$ and $\Lambda^i$ are specific functions of the transformation parameters.

\subsection{Lifshitz algebra}
\label{sec:Lifshitz}

In the previous sections, we saw that the presence of boosts complicates the situation considerably, since a number of structures transform in a noncovariant way under them. Here, we investigate another type of nonrelativistic spacetime symmetry, the Lifshitz algebra,  which can be obtained from the Galilei one by discarding the boosts. By doing so, the presence of the $U (1)$ symmetry associated with the central extension becomes unnecessary, since it decouples from the spacetime generators and turns into an internal symmetry.

Now all the structures can be classified in terms of irreducible representations of the $\mathfrak{so}(n-1)$ algebra of spatial rotations. The corresponding transformation properties of the fields can be read from the tables in Sec.~\ref{sec:Galilei}, as well as from Eqs.~\eqref{scaling_gauge} and~\eqref{scaling_strength}. For the Lifshitz algebra,  $n_{\m\n}$, $\t^{ij}_{\m\n}$, and $W_{\m\n}$ are identical to the ones in~\eqref{mod_field_str}, whereas the spatial torsion reads
\be
\begin{aligned}
\label{field_str_2}
e ^ i _ {\m \n} & =  2\p _ {[\m}e ^ i _ {\n]} + 2\t _ {[\m} ^ {i j} e _ {\n]  j} 
+2 W _{[ \m} e ^ i _ {\n]} \ .
\end{aligned}
\ee
Notice that all field strengths transform covariantly.

Imposing the following set of constraints,
\be
n _ {\m \n} E ^ \m _ i V ^ \n = 0\ , ~~~ e ^ i _ {\m \n}E ^ \m _ j E ^ \n _ k = 0 \ , ~~~e ^ {\l [i \r.} _ {\m \n} E ^ {\l. j \r ] \m} V ^ \n = 0 \ , ~~ 
e ^ {i} _ {\m \n} E ^ {\m} _ i V ^ \n = 0 \ ,
\ee
enables us to express in terms of the vielbein the connection that is once again given by~\eqref{spatial_theta_W}, and the Weyl gauge field whose spatial part is~\eqref{spatial_W}, whereas its temporal part reads
\be
W _ t=\frac{2}{n-1}\p_{[\m}e_{\n]}^{\l (i \r.}E^{\l.j \r ) \m}V^\n \ .
\ee
The above results are completely analogous to the ones in the torsionful relativistic theory~\cite{Karananas:2015eha}.

\section{Conclusions}
\label{sec:conclusions}

The aim of this paper was to clarify certain issues related to the gauging of nonrelativistic symmetries. After reviewing in detail the basic ingredients of the coset construction, we first presented a systematic way of building a locally invariant Galilei theory from a globally invariant one. 
We found that within this approach the term corresponding to the interaction of a spin $s$ and the magnetic field is automatically included. In this case, no modification of the transformation properties of the $U(1)$ gauge field is needed in order to achieve invariance of the action under boosts. 

We demonstrated how the covariant constraints can be used in order to eliminate redundant (unnecessary) degrees of freedom. It should be emphasized once again that it is not consistent to set to zero the spatial torsion, unless the temporal torsion vanishes as well (provided no Goldstone bosons are introduced). 

We then turned to the question of how the addition of dilatations changes the situation. We showed that there are no Weyl invariant theories with vanishing temporal torsion, i.e. with global time. The condition of temporal torsionlessness is not covariant under local scale transformations.
On the contrary, when torsion is present, it is always possible to express the spatial part of the Weyl gauge field in terms of geometric data.
We showed, however, that for general backgrounds it is not possible to eliminate the temporal part of the Weyl vector. 
Nevertheless, as we saw, it may happen that the aforementioned field does not appear in the action. As a result, invariance under Weyl rescalings does not necessarily require the introduction of $W_t$.
  
Finally, we discussed Lifshitz-invariant theories. In this case, the field strengths transform covariantly, since we relaxed the requirement of having invariance under Galilei boosts. In these theories both the temporal and the spatial parts of the Weyl field can always be expressed in terms of the vielbein. 

The fact that for the cases considered in the present paper the Weyl vector can be (partly) eliminated in favor of other degrees of freedom, should not come as a surprise. This is nothing else than torsion playing the role of the Weyl gauge field. It would be interesting to investigate the behavior of the propagating modes when terms bilinear in the various field strengths are taken into account. Similar analysis for relativistic theories has been carried out in~\cite{Karananas:2014pxa}, see also references therein.

\section*{Acknowledgements}
The work of G.K.K. and A.M. is supported by the Swiss National Science Foundation and by the Ambizione grant of the Swiss National Science Foundation.

\bibliographystyle{utphys}
\bibliography{NR_weyl_gauging_coset}{}

\end{document}